\documentclass[traditabstract]{aa}  

\usepackage{graphicx}
\usepackage{txfonts}
\begin{document}
   \title{Photometric properties of stellar populations in Galactic globular clusters: the r{\^o}le of the Mg-Al anticorrelation}

\authorrunning{Cassisi et al.}
\titlerunning{Mg-Al anticorrelation and photometric properties of globular cluster stars}

   \author{Cassisi, S.\inst{1}, Mucciarelli, A.\inst{2}, Pietrinferni, A.\inst{1}, Salaris, M.\inst{3}, \and Ferguson, J.\inst{4}}

   \institute{ INAF- Osservatorio Astronomico di Teramo, Via Mentore Maggini, 64100 Teramo, Italy\\  
              \email{cassisi,pietrinferni@oa-teramo.inaf.it}
         \and
            Dipartimento di Fisica e Astronomia, Universit{\'a} degli Studi di Bologna, Viale Berti Pichat, 6/2 - 40127, Bologna, Italy\\
            \email{alessio.mucciarelli2@unibo.it}
            \and            
             Astrophysics Research Institute, Liverpool John Moores University, Twelve Quays House, Birkenhead, CH41 1LD\\
             \email{ms@astro.livjm.ac.uk}
           \and            
              Department of Physics, Wichita State University Wichita, KS 67260, USA\\
               \email{jason.ferguson@wichita.edu}
}

   \date{Received xxx, xxx; accepted  xxx, xxx}

 \abstract{We have computed low-mass stellar models and synthetic spectra for an initial chemical 
composition that includes the full C-N, O-Na, and Mg-Al abundance anticorrelations observed in second generation stars belonging to  
a number of massive Galactic globular clusters.  
This investigation extends a previous study that has addressed the effect of only the C-N and O-Na 
anticorrelations, seen in all globulars observed to date. 

We find that the impact of Mg-Al abundance variations at fixed [Fe/H] and 
Helium abundance is negligible on stellar models and isochrones 
(from the main sequence to the tip of the red giant branch) and 
bolometric corrections, when compared to the effect of C-N and O-Na variations.

We identify a spectral feature at 490-520 nm, for low-mass main sequence stars, caused by 
MgH molecular bands. This feature has a vanishingly small effect on 
bolometric corrections for Johnson and Str\"omgren filters that cover that spectral range. 
However, specific narrow-band filters able to 
target this wavelength region can be powerful tools to investigate the Mg-poor unevolved stars and 
highlight possible splittings of the MS due to variations of Mg abundances.
} 

\maketitle

\section{Introduction}

Modern spectroscopic surveys of several Galactic globular clusters (GCs) 
have confirmed earlier results about the existence of 
well-defined chemical patterns in the atmospheres of  
stars within individual GCs, overimposed to the standard $\alpha$-enhanced metal mixture typical 
of field halo stars, as reviewed in, e.g., \cite{gcb:12}. 
More in detail, there is undisputed evidence that all GCs studied so far display  
anticorrelations between C and N, plus O and Na --  the so-called  {\sl C-N-O-Na anticorrelations} -- amongst 
their stars, at fixed [Fe/H] (but a handful of exceptions 
that show also a spread of [Fe/H]). These anticorrelations go in the sense of increasing N whilst decreasing C, and 
increasing Na while decreasing O,  
with respect to the baseline $\alpha$-enhanced metal composition. Many clusters 
-- but not all -- show also an anticorrelation between Mg and Al -- the 
{\sl Mg-Al anticorrelation}--  whereby also Mg is depleted and Al enhanced 
compared to the baseline $\alpha$-enhanced mixture, as discussed in \cite{gratton:01,carretta:09}
for several GCs.
The presence of these abundance patterns is a  
signature of matter processed during H-burning by high-temperature proton capture reactions, like the 
\emph{NeNa-} and \emph{MgAl-cycle}. Given that H-burning synthesizes He, one 
expects that these abundance patterns are associated with He-enhancement, 
as extensively discussed in, e.g., \cite{cs:13}.

Given that observations detect these abundance anticorrelations from the fainter portion of the main sequence (MS) -- 
as recently shown by \cite{milonevlm:12} --  all the way to the tip of the red giant branch (RGB), without signatures of dilution due to the 
varying size of the convective envelope, it is clear that their origin must be 
primordial\footnote{It is important to mention that \lq{overimposed}\rq\ to these primordial
abundance anticorrelations, there is also an evolutionary effect during the RGB phase, 
both in globular clusters and in the Galactic field, whereby the C abundance is 
increasingly depleted and N progressively enhanced 
after the RGB bump. This is usually interpreted as due to the occurrence
of additional mixing processes, as discussed in, e.g. \cite{jp:12} for the case of giant stars
in the GC M~13.} thus challenging the canonical idea that GCs 
host coeval stars, all formed with the same initial chemical composition. 

Additional, clear-cut evidence for multiple stellar populations in individual GCs  
has been provided by high-precision HST photometry, that has revealed the existence of multiple 
MSs  - - as in the cases of $\omega$ Cen (see, e.g. King et al. (2012) and references therein),
and NGC~2808 (Piotto et al. 2007) -- and/or multiple Sub-Giant Branch (SGB) - as for 
NGC~1851 (Milone et al. 2008), 47~Tuc (Milone et al. 2012), NGC~362, NGC~5286, NGC~6656, NGC~6715, NGC~7089
(Piotto et al. 2012), NGC~6388 and NGC~6441 (Bellini et al. 2013) -, and/or 
multiple RGB sequences -- as in the cases of NGC~6121 (Marino et al. 2008), and NGC~6656 (Marino et al. 2011) --,
in the colour magnitude diagrams (CMDs) of several GCs. 
The observed number of sequences and their photometric properties 
do change significantly from cluster to cluster, and strongly
depend on the adopted photometric system. Passbands including strong molecular bands 
of CN (bandheads at 388.3 and 421.6 nm), NH (around 345.0 nm) or CH (around 430.0 nm) 
-- see, e,g, \cite{grundahl:98}, \cite{grundahl:99}, \cite{carretta:11}, \cite{milone:12} and \cite{piotto:10} -- 
are particularly suited to disclose photometrically multiple populations.

The commonly accepted scenario envisages that in a GC (an) additional generation(s) of stars (second generation - SG) 
form(s) from the ejecta of intermediate-mass and/or massive stars, belonging to the first stellar population 
(first generation - FG) originated  
during the early phase of the cluster evolution. The chemical composition 
of these ejecta shows \lq{signatures}\rq\ of high-temperature proton captures, and 
after some dilution with pristine (unpolluted) matter -- that seems   
necessary to explain the observations, as discussed in \cite{dercole:11} -- SG stars would be then 
formed from gas characterized by the observed light-element anticorrelations and some level of He enhancement. 

The existence of a strong correlation between spectroscopic signatures of the distinct subpopulations and their 
distribution along the observed CMDs, has suggested that the chemical patterns of SG stars do affect both the 
evolutionary properties of these stars and their spectral energy distribution.
To this purpose, several theoretical investigations -- see, e.g., \cite{salaris:06,cassisi:08,ventura:09,pietrinferni:09} -- 
have studied  
the effect of SG chemical patterns on the structure and evolution of low-mass, low-metallicity stars. 
Recently \cite{sbordone:11} have also performed the first detailed investigation 
of the effect of SG chemical compositions on spectra,  bolometric corrections, and colours, 
for widely used blue to 
near-infrared photometric filters. 
These bolometric corrections and colour-transformations, coupled to theoretical 
isochrones with the appropriate chemical composition, represent the first complete and self-consistent set of
theoretical predictions for the effect of FG and SG stars on the observed cluster CMDs.

The main outcomes of these theoretical analyses can be summarized as follows. Colour and magnitude changes are 
largest in blue photometric filters, bluer than the standard {\sl B} one, independently of using 
broad- or intermediate bandpasses such as the Str{\"o}mgren filters,
so providing a sound interpretation of the observational findings previously mentioned. In particular, 
CMDs involving $uvy$ and {\it UB} filters  are best suited to detect photometrically the presence of multiple
stellar generations in individual clusters. The reason is that these filters are particularly 
sensitive to the N abundance (via its direct effect on the strength of the CN and NH molecular bands),
with the largest variations affecting the RGB and lower MS.  The optical bands  
BVI are expected to display multiple sequences only if the different
populations are characterized by variations of the C+N+O sum and/or He abundance, that lead to changes in the bolometric 
luminosity and effective temperature of stellar models, but leave the flux distribution above 400~nm  practically unaffected. 
These results have been also confirmed and extended to the ultraviolet and infrared WFC3 photometric system  
by \cite{milone:12} and \cite{milonevlm:12}, respectively. 

As already mentioned, the extensive and accurate spectroscopical survey performed 
by \cite{bragaglia:10,carretta:09} has shown that the {\sl Mg-Al anticorrelation} 
is present only in a subsample of clusters; in particular, extreme Mg depletions and Al enhancements have been 
detected only in very massive clusters like NGC2808, NGC6388 and NGC6441, and/or metal-poor ones, like NGC6752. 
The most extreme values have been found in NGC2808, that with its triple 
MS  -- explained by three distinct initial He abundances --  
and its very extended Na-O 
ant-correlation, is one of the most peculiar GCs in terms of the hosted populations. 
Finally, we mention the discovery by \cite{mucciarelli:12} 
of a large sample of very Mg-poor stars in NGC~2419, not coupled to a large 
Al-enhancement, that could be a unique case of a stellar population directly 
formed from the ejecta of asymptotic and super asymptotic giant branch stars, as discussed by \cite{ventura:12}.

To date, the effect of the {\sl Mg-Al anticorrelation} on the CMDs of Galactic GCs has been however largely unexplored
The impact of selectively enhancing the abundance of Mg with respect to the other 
heavy elements in the stellar spectra and/or models 
has been investigated theoretically to different degrees by \cite{cassisi:04, dotter:07, lee:09, vandenberg:12}. 
To the best of our knowledge, there has been no attempt to assess the effect on both stellar models and spectra  
of a change in the relative abundances of both Mg and Al, consistent with the spectroscopic measurements. 

The aim of this work is therefore to study the possible impact of a {\sl Mg-Al anticorrelation} 
added to {\sl C-N-O-Na anticorrelations}, on the CMDs of 
GCs, by computing stellar models and bolometric corrections 
that account self-consistently for realistic FG and SG chemical patterns.
The results of this investigation will help studies aimed at tracing the presence of the
distinct populations in a cluster, based on pure 
photometric analyses along all the relevant evolutionary stages -- see, e.g., \cite{monelli:13} -- 
given that high-resolution spectroscopic analyses 
are so far limited to a relatively small number of objects, mostly bright RGB stars .

The structure of the paper is as follows: the next section presents the stellar evolution models, whilst Section~3 
describes the model atmosphere and synthetic spectrum calculations, followed by our conclusions.

\begin{figure*}
\centering
\vskip -4cm
\includegraphics[scale=0.9]{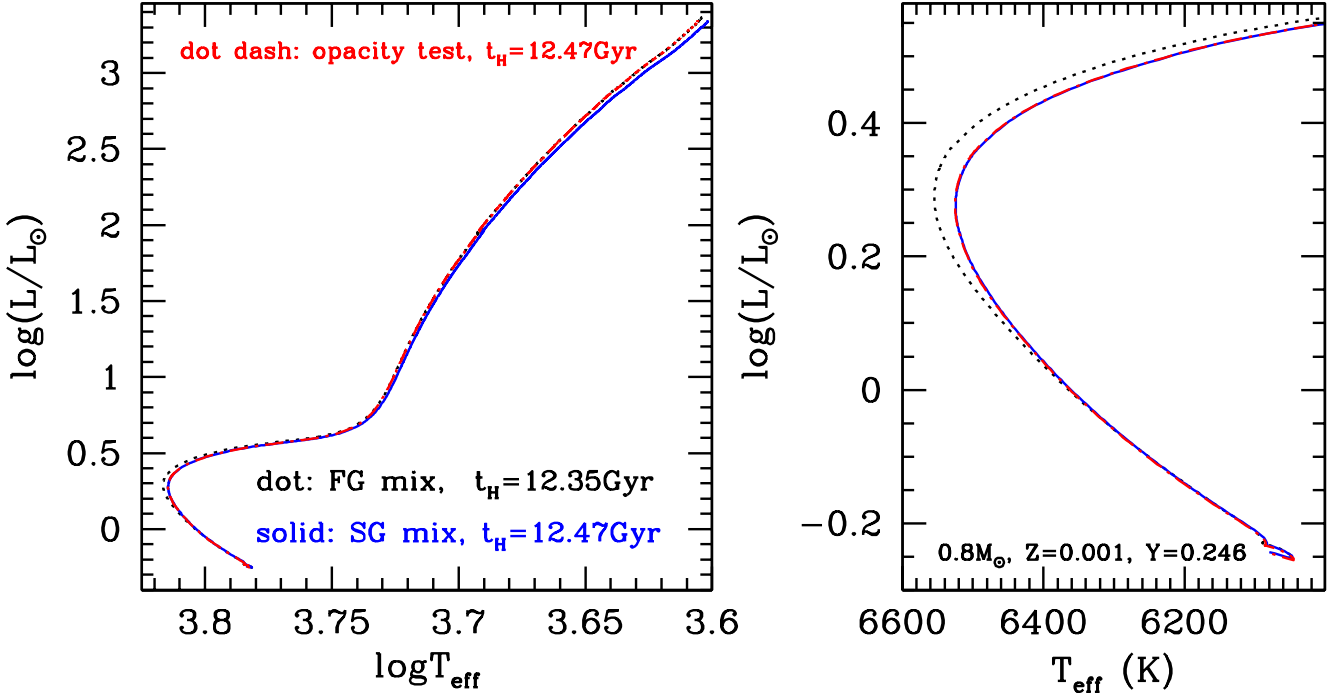}
\vskip -4cm
\caption{\textit{Left panel}: evolutionary tracks for ${\rm 0.8M_\odot}$ models and Z=0.001, Y=0.246 FG and SG chemical compositions. 
An additional numerical experiment with modified opacity tables is also shown (see text for details).  MS lifetimes are listed. 
\textit{Right panel}: as left panel but with an enlargement of the MS-TO-SGB region is shown.}
\label{fig:trk}
\end{figure*}

\section{The evolutionary framework}

We consider here a reference BaSTI\footnote{http://www.oa-teramo.inaf.it/BASTI} $\alpha-$enhanced 
isochrone from \cite{pcsc:06} with an age of 12~Gyr, initial He abundance and metallicity equal to Y=0.246 and Z=0.001, 
respectively. 
For the $\alpha-$enhanced heavy element mixture of the BaSTI database, these choices correspond to a ${\rm [Fe/H]=-1.62}$. This isochrone 
is representative of FG stars present in a typical Galactic GC. 

\begin{figure}
\centering
\includegraphics[scale=0.45]{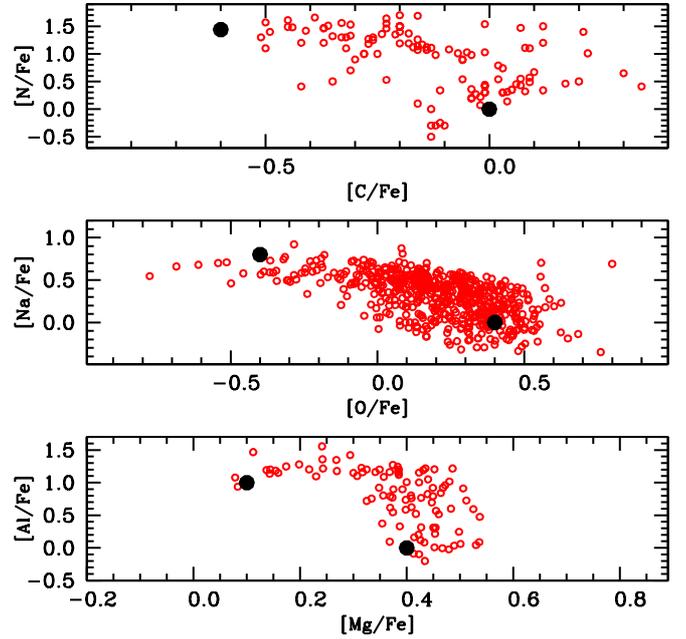}
\caption{The three panels display a comparison of the abundances of  
our FG and "CNONa2MgAl" compositions (filled large circles) with selected observational data for -- from top to bottom --  
the [C/Fe]-[N/Fe] diagram (data for 47~Tuc and NGC6752 from Briley et al.~2004 and Carretta et al.~2005), 
[O/Fe]-[Na/Fe] diagram (data for several clusters from Carretta et al. 2009a), [Mg/Fe]-[Al/Fe] diagram 
(data for NGC6752 from Carretta et al.~2012).
}
\label{obspatt}
\end{figure}

As for the SG composition, given that in \cite{pietrinferni:09} and \cite{sbordone:11} (see their Table~2) 
we have already explored the impact on both isochrones and spectra 
of He-enhancement and {\sl C-N-O-Na anticorrelations} 
both with and without keeping constant the sum of CNO elements, we now focus our attention on the 
presence of an additional {\sl Mg-Al anticorrelation} keeping the ${\rm C+N+O}$ total abundance constant. 
In more detail, we have selected one of the three metal mixtures used in \cite{sbordone:11},  
the so-called "CNONa2" mixture characterized 
by a depletion of C and O by 0.6~dex and 0.8~dex by mass, respectively, and by enhancements of N and Na by 
1.44~dex and 0.8~dex, compared to the FG metal abundances. To these changes, we have also added an enhancement of Al 
by 1~dex, and a depletion of Mg by 0.3~dex, with respect to the FG mixture. 
The Mg-Al variations are 
similar to the most extreme cases, observed in clusters like NGC~6752 and NGC~2808 
(see Carretta et al.~2009b, 2012).
Our adopted CNONaMgAl 
abundance pattern corresponds to [(C+N)/Fe]=0.73 (compared to [(C+N)/Fe]=0.0 for the 
FG $\alpha$-enhanced mixture), [(C+N+O)/Fe]=0.37 (within 0.5\% of the FG value), 
[(Mg+Al)/Fe]=0.28 (compared to [(Mg+Al)/Fe]=0.38 for the FG mixture). 
We denote this mixture as "CNONa2MgAl", that corresponds to extreme values of the light element  
anticorrelations observed in Galactic GCs. 
In this way we maximize the impact of the SG composition on both models and spectra. 
The chemical abundances for both the SG "CNONa2MgAl" and FG metal mixtures    
are listed in Table~\ref{tab:chim}.
Figure~\ref{obspatt} compares 
the values of [C/Fe], [N/Fe], [O/Fe], [Na/Fe], [Mg/Fe], [Al/Fe] for our FG mixture and 
the "CNONa2MgAl" composition, with observations of a few clusters.

We have then computed low-mass stellar models with ${\rm [Fe/H]=-1.62}$ and Y=0.246, for both metal compositions 
(that correspond to the same total metallicity Z=0.001). 

\begin{table}[ht]
\caption{Mass and number fractions (normalized to unity) for the 
FG and the "CNONa2MgAl" metal mixtures. }
\label{tab:chim}     
\centering
{\scriptsize                          
\begin{tabular}{lcccc} 
\hline
   & \multicolumn{2}{c}{First Generation}        & \multicolumn{2}{c}{CNONa2MgAl}     \\
   & Number frac.  & Mass frac.  &  Number frac.  & Mass frac.   \\ 
\hline
C   & 0.108211 & 0.076451 &  0.02738 &  0.01935  \\
N   & 0.028462 & 0.023450 &  0.71229 &  0.64723  \\
O   & 0.714945 & 0.672836 &  0.11404 &  0.10766  \\
Ne & 0.071502 & 0.084869 & 0.07197 &  0.08555  \\
Na & 0.000652 & 0.000882 &  4.13689E-3 & 5.60988E-3  \\
Mg & 0.029125 & 0.041639 &  0.01466 &  0.02099  \\
Al  & 0.000900 & 0.001428 &   9.05888E-3 &  0.01440  \\
Si  & 0.021591 & 0.035669 &   0.02173 &  0.03596  \\
P   & 0.000086 & 0.000157 &  8.65627E-5 & 1.58266E-4  \\
S   & 0.010575 & 0.019942 & 0.01064 & 0.02010  \\
Cl  & 0.000096 & 0.000201 &  9.66281E-5 &  2.02621E-4  \\
Ar  & 0.001010 & 0.002373 &   1.01661E-3 & 2.39214E-3  \\
K   & 0.000040 & 0.000092 &  4.02617E-5  & 9.27419E-5 \\
Ca & 0.002210 & 0.005209 & 2.22446E-3 &  5.25101E-3   \\
Ti   & 0.000137 & 0.000387 & 1.37896E-4  & 3.90121E-4  \\
Cr  & 0.000145 & 0.000443 &  1.45949E-4 & 4.46573E-4  \\
Mn & 0.000075 & 0.000242 & 7.54907E-5 & 2.43952E-4   \\
Fe  & 0.009642 & 0.031675 &  9.70508E-3 & 0.03193 \\
Ni  & 0.000595 & 0.002056 &  5.98893E-4 & 2.07258E-3  \\
\hline
\end{tabular}}
\end{table}

Opacities for the stellar interiors have been obtained from 
the Livermore Laboratory OPAL opacity website
\footnote{http://opalopacity.llnl.gov}, while the low-temperature opacity tables have been computed 
for this specific project. The other physics inputs are as described by \cite{pcsc:06} for the FG models.

Figure \ref{fig:trk} displays the theoretical Hertzsprung-Russell (H-R) diagram of 
${\rm 0.8M_\odot}$  models calculated with the CNONa2MgAl and FG mixtures, respectively, from the MS to the tip of the RGB. 
The two tracks overlap almost perfectly along the 
MS, and the turn off (TO) luminosities differ by only ${\rm \Delta\log(L/L_\odot)\approx0.017}$, the FG track being brighter. 
At the TO ${\rm T_{eff}}$ differences are $\sim30$~K, whilst at the base of the RGB 
is almost zero, increasing to just $\sim20$~K at the tip of the RGB, FG models being hotter. 
Lifetimes are also almost identical, 
the "CNONa2MgAl" model being only $\sim1$\% older at the TO.
Core masses at the He-flash, as well as RGB bump luminosity and envelope He mass fraction after the first dredge up are 
also the same.
The (very) small differences observed around the TO and the 
SGB are due to the fact that, as previously mentioned, the CNONa2MgAl mixture
is slightly more CNO-rich (by $\sim$0.5\%) than the mixture adopted for the FG.

\begin{figure}
\centering
\includegraphics[scale=0.45]{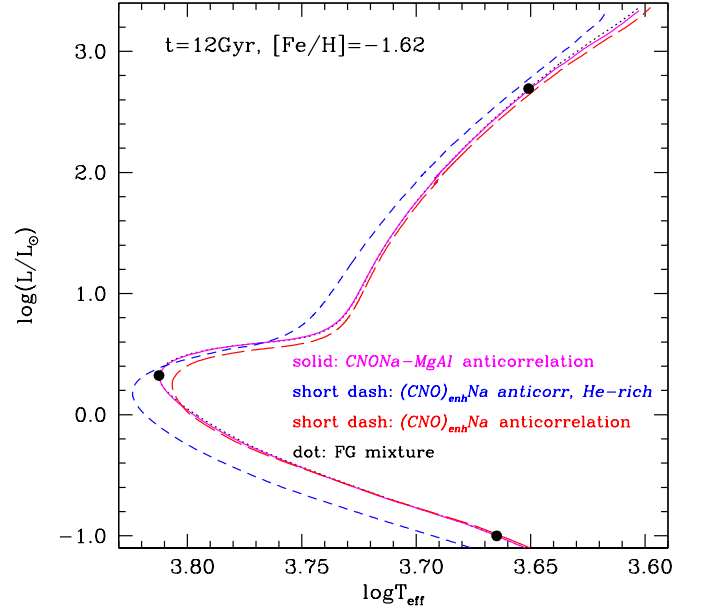}
\caption{Isochrones for an age of 12~Gyr and ${\rm [Fe/H]=-1.61}$, under various assumptions about the metal mixture and the initial He 
abundance. The {\it CNONa-MgAl} label denotes the mixture with the C-N, Na-O and Mg-Al anticorrelations, and Y=0.246; 
$(CNO)_{enh}Na$ denotes a mixture with C-N, Na-O anticorrelation, and the CNO sum enhanced by a factor of two with respect  
to the FG mixture -- as in \cite{sbordone:11} -- and Y=0.246;  $(CNO)_{enh}Na, He-rich$ denotes 
the same mixture as the $(CNO)_{enh}Na$ case, 
but with Y=0.40. Filled circles mark selected points
along the isochrones for which we show in the following the corresponding synthetic spectra.}
\label{fig:iso_anticor}
\end{figure}

To highlight the effect on the evolutionary tracks 
of high- and low-temperature opacities with the {\sl Mg-Al anticorrelation}, 
we have calculated additional evolutionary models, also displayed in Fig.\ref{fig:trk}. 
For this calculation we have included the "CNONa2MgAl" metal mixture 
in both nuclear network (where the effect is zero, because 
neither Mg nor Al are involved in H-burning reactions during the MS for low-mass stars) 
and high-temperature opacities (i.e. ${\rm \log{T}>4.2}$), 
and the FG mixture for the low-temperature opacities. The right panel of Fig.~\ref{fig:trk} shows clearly 
that around the TO the opacities of the interiors determine 
the location of the track (this new track overlaps at the TO with the complete calculation for  
the "CNONa2MgAl" mixture), whilst the low-temperature opacities fix the 
${\rm T_{eff}}$ of the RGB (the track overlaps with the FG track along the RGB) 
in agreement with results by \cite{salaris:93} and \cite{cassisi:05}.

\begin{figure}
\centering
\includegraphics[scale=0.45]{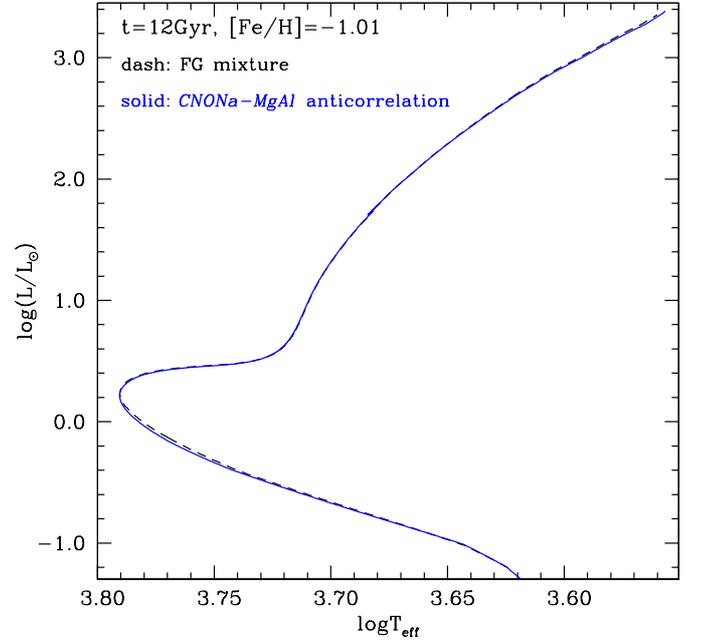}
\caption{Isochrones for an age of 12~Gyr and ${\rm [Fe/H]=-1.01}$, Y=0.248, computed for both FG (dashed line) 
and "CNONa2MgAl" (solid line) chemical compositions.}
\label{fig:isoz43}
\end{figure}

With the help of additional tracks for different masses we have calculated 
a typical GC isochrone for 12 Gyr, [Fe/H]=$-$1.62, Y=0.246, from the MS (minimum mass equal to 0.5${\rm M_{\odot}}$) 
to the tip of the RGB, and the "CNONa2MgAl" mixture.
This is compared in Fig.~\ref{fig:iso_anticor} with additional isochrones for the same age and the 
same [Fe/H]=$-$1.61, that we calculated in \cite{sbordone:11}.
First of all, the 
"CNONa2MgAl" isochrone overlaps with the FG isochrone and the isochrone for the "CNONa2" mixture 
(they are indistinguishable from one another). 
The small differences between the 0.8${\rm M_{\odot}}$ tracks for FG and "CNONa2MgAl" mixtures    
are erased because of the slightly different evolutionary timescales that 
produce a slightly different mass distribution along the 
corresponding isochrones. The perfect agreement between FG and  "CNONa2MgAl" isochrones is confirmed also at an higher metallicity, 
e.g. [Fe/H]=$-$1.01 (Y=0.248), as displayed by  
Fig.~\ref{fig:isoz43}.
For the sake of comparison we display in  Fig.~\ref{fig:iso_anticor} 
also [Fe/H]=$-$1.61 isochrones 
with extreme values of the {\sl C-N-O-Na anticorrelations} and the CNO sum enhanced by a factor 
of two -- mixture denoted as "CNONa1" in \cite{sbordone:11} -- and the same 
CNO-enhanced mixture with also an enhanced He-abundance (Y=0.40). 

As a consequence of these evolutionary computations, we can extend the analysis of the impact of light element 
anticorrelations on the morphology of isochrones in the H-R diagram performed by \cite{pietrinferni:09} and 
\cite{sbordone:11}. 
As long as the sum of the CNO elements and the initial He abundance is unchanged, 
the presence of C-N, O-Na and Mg-Al anticorrelations does not affect the H-R diagram of theoretical isochrones 
of fixed age and [Fe/H]. An increase of He (at fixed metal mixture) shifts the whole isochrone towards hotter ${\rm T_{eff}}$, 
and produces a slightly fainter TO, whereas an increase of the CNO sum makes the TO and SGB fainter, leaving the rest of the isochrone 
basically unchanged.

\begin{table}
\caption{Selected points along the "CNONa2MgAl" isochrone, chosen for the model atmosphere computations.}         
\label{tab:keypoint}     
\centering                          
\begin{tabular}{cc}       
\hline                
${\rm T_{eff}}$ & $\log g$     \\    
(K) & ($\mathrm{cm/s^2}$)              \\
\hline 
4100  & 0.50  \\
4476  & 1.20  \\
4892  & 2.06  \\
5312  & 3.21  \\
5854  & 3.78  \\
6490  & 4.22  \\
6131  & 4.50  \\
4621  & 4.77  \\
\hline
\end{tabular}
\end{table}

\section{Synthetic spectra}

As a next step, we have investigated the effect of the Mg-Al anticorrelation on the spectral energy 
distribution of low-mass stars.
Given that the impact of a mixture with extreme values of the C-N and O-Na anticorrelations 
has been already investigated by \cite{sbordone:11}, here we study whether the effect of an additional 
(extreme) Mg-Al anticorrelation is appreciable compared to a mixture with only C-N and O-Na extreme 
anticorrelations.  

To this purpose we have selected key-points -- listed in Table~\ref{tab:keypoint} -- along the 12~Gyr, 
[Fe/H]=$-$1.62 FG isochrone of Fig.~\ref{fig:iso_anticor}, that match those chosen by \cite{sbordone:11}.
For each of these (${\rm T_{eff}}$, log($g$)) points we have computed 
model atmospheres plus the corresponding synthetic spectra for two 
{\bf assumptions} about the heavy element distributions, and calculated 
bolometric corrections for both the Johnson-Cousins {\sl UBVI} 
and the Str{\"o}mgren {\it uvby} photometric systems.

Model atmospheres 
have been calculated as in \cite{sbordone:11}, by employing 
the ATLAS~12 code, described by \cite{castelli:05} and \cite{sbordone05}, 
in its last version as available at the website of F. Castelli
\footnote{http://wwwuser.oat.ts.astro.it/castelli/sources/atlas12/}.
At variance with the ATLAS~9 code -- widely employed because of 
its computational speed -- that uses pre-tabulated line opacities in the form of 
Opacity Distribution Functions (ODFs),   
ATLAS~12 employs the opacity sampling method for the line opacity computation. 
This approach (even if computationally more expensive) enables to calculate easily 
models with arbitrary chemical compositions, whilst ATLAS~9 is tied to the 
chemical mixture adopted in the ODFs. 
All ATLAS~12 models assume plane-parallel geometry and LTE for all chemical 
species.

\begin{figure}
\centering
\includegraphics[scale=0.45]{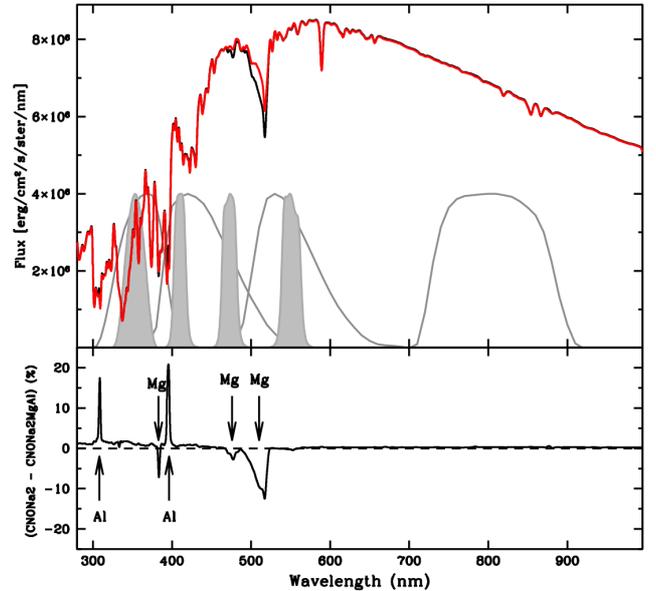}
\caption{{\textit Top panel}: Flux distribution for the MS model with ${\rm T_{eff}=4621}$~K and 
${\rm \log(g)=4.77}$ calculated with "CNONa2" (black line) and the 
"CNONa2MgAl" (red line) metal compositions. 
The transmission curves for the Johnson-Cousins {\sl UBVI} (thin black lines) and the Str{\"o}mgren 
{\it uvby} filters (grey-shaded regions) are also shown. {\textit Bottom panel}: Relative flux difference between 
the two spectral energy distributions, as a function of the wavelength.}
\label{fig:spectrum_ms}
\end{figure}

\begin{figure}
\centering
\includegraphics[scale=0.45]{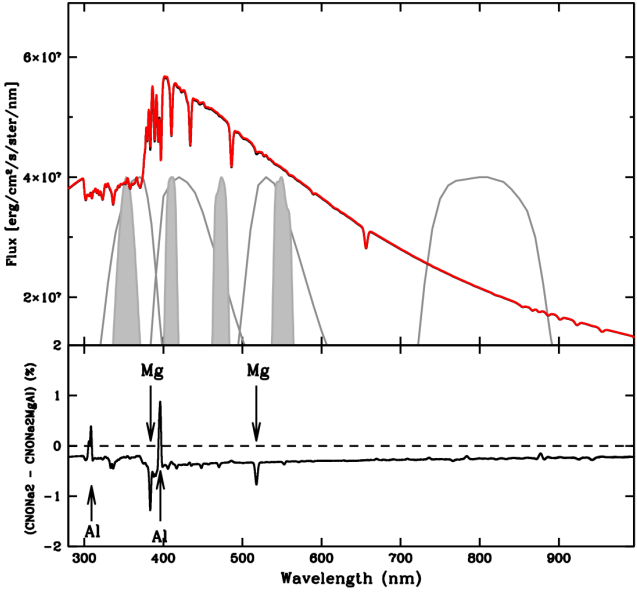}
\caption{As in Fig.~\ref{fig:spectrum_ms} but for the TO model with ${\rm T_{eff}=6490}$~K and 
${\rm \log(g)=4.22}$. }
\label{fig:spectrum_to}
\end{figure}

For each key point, we calculated two model atmospheres, adopting different 
chemical compositions: 
{\sl (a)}~the SG chemical mixture with only C-N and O-Na anticorrelations, labelled "CNONa2" by \cite{sbordone:11}; 
{\sl (b)}~the SG "CNONa2MgAl" metal mixture that includes also the Mg-Al anticorrelation.

We note that all the model atmospheres, consistently with the evolutionary models, have been computed 
by adopting ${\rm [Fe/H]=-1.62}$~dex, Y=0.246. We also adopted 
a microturbulent velocity equal to 2 km/s, as for the main set of calculations by \cite{sbordone:11}.

Synthetic spectra have been calculated with the code SYNTHE \cite{sbordone05} 
in the wavelength range from 200 to 1000 nm.
We have included all atomic and molecular 
transitions available in the last version of the Kurucz/Castelli database
\footnote{http://wwwuser.oat.ts.astro.it/castelli/linelists.html}, excluding 
only TiO lines, because in the investigated cases there are no prominent TiO
bands (and the inclusion of these transitions increases significantly the computational time).
Following the procedure already adopted in \cite{sbordone:11}, the 
spectra have been calculated at high resolution and then convolved  
with a Gaussian profile with FWHM=~1700 km/s (corresponding to a 
spectral resolution $\lambda/\Delta \lambda \sim$175).

\begin{figure}
\centering
\includegraphics[scale=0.45]{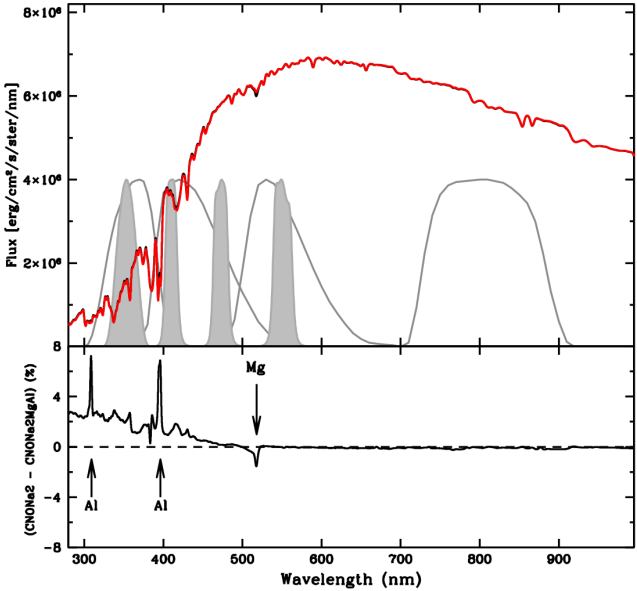}
\caption{As in Fig.~\ref{fig:spectrum_ms} but for the RGB model with ${\rm T_{eff}=4476}$~K and ${\rm \log(g)=1.2}$. }
\label{fig:spectrum_rgb}
\end{figure}

Figures~\ref{fig:spectrum_ms}, 
~\ref{fig:spectrum_to} and ~\ref{fig:spectrum_rgb} compare synthetic spectra 
for the two compositions, and the points displayed as filled circles along the FG isochrone 
in Fig.~\ref{fig:iso_anticor}.

The largest differences between the displayed spectra 
with the "CNONa2" and "CNONa2MgAl" mixture are found
for the MS star.
The prominent feature detectable between $\sim$495 and $\sim$521 nm corresponds to the (0,0) vibrational 
band of the MgH $A^{2}\Pi$-$X^{2}\Sigma$ system  
plus the prominent lines of the Mg {\sl b} triplet. A weaker Mg feature is  visible at $\sim$475 nm 
and corresponds to the (1,0) band of MgH.
At bluer wavelengths, the absorption due to the Mg triplet at $\sim$383.8 nm is also visible.  
Concerning Al, the main variations between the two synthetic spectra are  close to the H and K Ca lines, 
due to the Al resonance lines at 394.4  and 396.1 nm, and around 309 nm (Al lines at 308.2 and 309.2 nm).

The same features can be recognized also in the synthetic spectra of the RGB star. In this case, 
we clearly observe the same Al lines, while the Mg features are less prominent, because the MgH bands weakens 
decreasing  the surface gravity (the variation at $\sim$520 nm is basically due to the 
Mg b triplet).

Concerning the comparison of the synthetic spectra for the TO star, the two spectra are basically 
indistinguishable: the main variations do not 
exceed $\sim\pm$1.5\% of the flux. Also in this case, these differences involve the Mg and Al transitions 
already mentioned for the two (${\rm T_{eff}}$, log($g$)) pairs discussed above.

\begin{figure}
\centering
\includegraphics[scale=0.45]{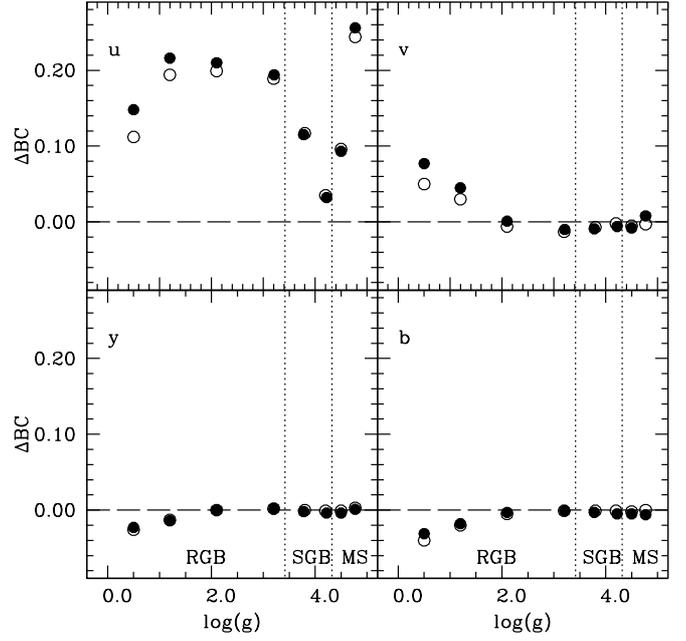}
\caption{Difference $\Delta$BC between the $uvby$ bolometric corrections  
for a standard $\alpha$-enhanced FG mixture and, respectively, the "CNONa2" SG mixture 
(open circles) and the "CNONa2MgAl" SG mixture, for the selected points of 
Table~\ref{tab:keypoint} (see text for details).}
\label{fig:deltabc1}
\end{figure}

\begin{figure}
\centering
\includegraphics[scale=0.45]{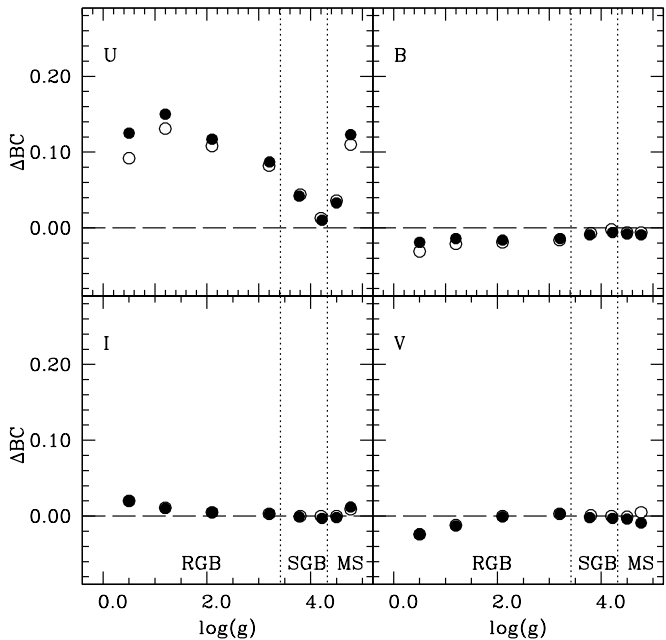}
\caption{As in Fig.~\ref{fig:deltabc1}, but for the {\sl UBVI} filters.}
\label{fig:deltabc2}
\end{figure}

We have then calculated for each spectrum bolometric corrections (BCs) 
following the methods by \cite{gir:02}, and 
the differences 
$\Delta$BC at each of the points listed in Table~\ref{tab:keypoint}, between the "CNONa2MgAl" and 
"CNONa2" spectra, for the {\sl UBVI} and $uvby$ photometric filters \footnote{As in 
\cite{sbordone:11} we employ  
the passband definitions provided by \cite{bessell:90} and \cite{stromgren:56}.}. 
These differences for each selected point have been then added to the difference of bolometric corrections between the 
FG $\alpha$-enhanced composition and the"CNONa2" metal distribution, as 
derived by \cite{sbordone:11} with the same codes employed in our analysis.
The resulting set of $\Delta$BC values include the effect of the full 
{\sl C-N-O-Na-Mg-Al anticorrelations}, and is compared to the case of only {\sl C-N-O-Na anticorrelations} 
in Figs.~\ref{fig:deltabc1} and \ref{fig:deltabc2} .

As can be easily 
gathered by comparing the filled and open circles in these two figures -- and   
as expected from the previous 
analysis of individual spectra -- the effect of the {\sl Mg-Al anticorrelation} 
is extremely small, almost always below $\sim$0.015~mag.
Only the two lowest gravity points along the upper RGB display 
differences up to 0.03-0.04~mag in the {\sl U}, $u$ and $v$ filters. 
In case of the point at log($g$)=1.20, the differences in {\sl U} are a factor $\sim$3 smaller than 
the corresponding difference between FG and "CNONa2" determined by \cite{sbordone:11}. For the $u$ 
filter differences are a factor $\sim$10 smaller. At log($g$)=0.5 the differences in {\sl U} and $u$ 
are a factor $\sim$3-4 smaller; only for the $v$ filter the difference is more comparable 
to the relatively small difference between FG and "CNONa2" BCs.

As a conclusion, the role played by the {\sl Mg-Al  
anticorrelation} in both the theoretical isochrones and bolometric corrections 
appears generally negligible compared to the effect of the {\sl C-N-O-Na   
anticorrelation}, at least for the photometric systems investigated here.

\section{Conclusions}

We have explored self-consistently the impact of a metal mixture that reflects extreme values of the 
C-N, O-Na, and Mg-Al anticorrelations observed in Galactic GCs, on both theoretical H-R diagrams and CMDs  
of old metal-poor populations, from the MS to the tip of the RGB. 
The main conclusion of our investigation is that the effect of the {\sl Mg-Al anticorrelation} on the evolutionary
properties of low-mass stars, and hence theoretical isochrones for old stellar populations,  is negligible. 
This confirms the theoretical inferences discussed by \cite{salaris:06}.
As for the BC scale, the effect is vanishing when compared to the r{\^o}le 
played by the {\sl C-N-O-Na anticorrelations}, but the case of BCs for the $v$ Str\"omgren filter, and only for stars near the tip of the RGB.  

Based on the isochrones displayed in Fig.~\ref{fig:iso_anticor}, the results about the BC scales 
by \cite{sbordone:11}, and Figs.~\ref{fig:deltabc1}, \ref{fig:deltabc2}, one can confirm the following general conclusions about 
the photometric properties of GC multiple populations, from the MS to the tip of the RGB:

\begin{itemize}

\item{Visible to near-infrared CMDs (e.g. ${\sl BVI}$ or Str\"omgren $y$) 
are generally unaffected by the presence of SG stars, unless this latter component has a different 
initial He content. In this case multiple sequences appear along the MS. This explains very nicely. i.e., 
the multiple MS of NGC2808 in optical CMDs, discussed by \cite{piotto:07}. Also a variation of the CNO sum 
affects optical CMDs, by altering the location of TO and SGB. This is a possible explanation for the double SGB of NGC1851 
in optical filters, discovered by \cite{milone:08} (see Cassisi et al.~2008), as well as for the double SGB
observed in NGC6656 (Marino et al.~2012). Both He and CNO affect optical CMDs because they 
hugely affect the stellar evolutionary properties, and hence modify the isochrone location in the  H-R diagram.}

\item{At shorter wavelengths {\sl C-N-O-Na-Mg-Al anticorrelations} affect the BC scale, so that even at constant CNO and He 
abundances, SG and FG stars will be 
distributed -- continuously, in case of a continuous distribution of the abundance anticorrelations -- 
amongst several separate sequences.
The quantitative effect will depend on the filter combination 
employed and the exact over- and underabundances with respect to the baseline FG composition. Variations of He and CNO add an 
additional level of separation in the CMDs, due to their effect on the underlying isochrones (both He and CNO sum) and BCs (mainly due 
to the CNONa anticorrelations).}

\end{itemize}

The general conclusion is that therefore anticorrelated 
Mg and Al variations do not leave any signature in CMDs of GCs, at least for the widely used filters 
discussed in this paper. However, we wish to notice that the most relevant 
difference between synthetic spectra of CNONa2 and CNONa2MgAl is observed in the MS spectrum  
shown in Fig.~\ref{fig:spectrum_ms} and is related to the MgH molecular bands. 
The standard filters investigated here are not able to isolate the effect of this molecular band 
at 490-520 nm, hence this feature has a vanishingly small effect on the corresponding BCs.
This feature can however be potentially useful 
to detect stars characterized by Mg depletion along the MS.
In general, the Mg depletion observed in GC stars is smaller than the O depletion,  
and stars with sub-solar [Mg/Fe] ratios are less common. 
Specific narrow-band filters centred on this spectral region can be powerful tools to investigate the Mg-poor unevolved stars and 
highlight possible new splittings of the MS due to relatively small variations of Mg.

\begin{acknowledgements}

We warmly thank our referee for her/his constructive comments. SC is grateful for financial support 
from PRIN-INAF 2011 "Multiple Populations in Globular Clusters: their 
role in the Galaxy assembly" (PI: E. Carretta), and from PRIN MIUR 2010-2011, 
project \lq{The Chemical and Dynamical Evolution of the Milky Way and Local Group Galaxies}\rq, prot. 2010LY5N2T (PI: F. Matteucci).
      
\end{acknowledgements}

\end{document}